# Reentrant cluster glass behavior in $La_2CoMnO_6$ nanoparticles

**J. Krishna murthy**[1] **and A. Venimadhav**[2,a]

[1]*Cryogenic Engineering Centre, Indian Institute of Technology, Kharagpur-721302, India*

## Abstract

Magnetic study on $La_2CoMnO_6$ nanoparticles revealed multiple magnetic transitions at 218 K, 135 K and below 38 K and the nature of the low temperature transition was unclear [J. Appl. Phys. 111, 024102 (2012)]. Presence of mixed valance states of Co and Mn has been confirmed from the XPS measurement and its presence along with antisite disorder affects in reducing the saturation magnetization of the nanoparticles. The zero-field–cooled and field-cooled bifurcation in dc magnetization, relaxation in zero-field–cooled magnetization and large enhancement in coercive field below the glassy temperature has been discussed. Frequency dependence of ac susceptibility using power law has revealed cluster glass behaviour. Further, the dc field superimposed on ac susceptibility and absence of memory effect in ac susceptibility has suggested the existence of non-interacting clusters comprising of competing interactions below 38 K. Competing magnetic interactions due to the presence of mixed valances and antisite disorder found to establish a reentered cluster glassy state in the nanoparticles.

a) Corresponding author: venimadhav@hijli.iitkgp.ernet.in

# 1. Introduction:

The spin glass (SG) materials are classified with localized magnetic moments whose interactions are characterized by quenched randomness. Strength of the competing exchange interactions and the disorder decides the nature of the glassy state to be in SG or cluster glass (CG) state.[1-4] In classical SG systems strong competition between FM and AFM interactions destroy the long-range magnetic order and leaves the system in a random spin configuration of frozen state.[1,2,4] The SG state has been observed in spinels, amorphous materials and largely in diluted magnetic alloys.[5-7] If the competing interactions are weak (either FM or AFM phase dominates) such materials show CG behavior with short range magnetic order. A good number of magnetic oxides particularly manganites and cobaltates show the CG nature.[8-10] Poddar et al [3] reported that in double perovskite $Sr_2Fe_{1-x}Mn_xMoO_6$ system, the observed CG behavior was attributed to the local spin frustration developed due to the competing interactions among the nearest neighbor and next nearest neighbors. On the other hand, a high temperature long range magnetic ordering can get partially or fully destroyed due to moderate competing interactions (arising from the magnetic disorder) that introduce some frustration among the set of spins that leads to a SG-like state at low temperatures known as RSG behavior.[11] Reentrant magnetic behavior has been found in a variety of disordered magnetic materials.[11,12,13]

In perovskite materials, structures and magnetic properties are very sensitive to the valance state of the transition metal ions and the doping at cation sites. Further, the size reduction of particles to nanoscale would break the long range magnetic exchange bonds and the translation symmetry of the lattice that pray to distinct magnetic interactions leading to glassy magnetic properties. SG state has been widely observed in a number of single[14] and double perovskite[15] compounds but only a few systems showed RSG nature.[11,16] Recently, antisite defect induced RSG has been noticed in a partially disordered $La_2NiMnO_6$ (LNMO) bulk system.[16] However, the high temperature FM ordering is completely broken and the systems prefers SG state in the LNMO nanoparticles.[17] On the other hand, isostructural $La_2CoMnO_6$ (LCMO) nanoparticles retains the high temperature FM transitions (one at 218 K and other at 135 K) in contrast to LNMO nanoparticles, however, another magnetic transition was observed below 38 K and the nature of this transition is unclear.[18] In this paper we report detailed magnetic study on LCMO nanoparticles using dc and ac magnetization techniques and show that the observed low temperature transition below 38 K has cluster type reentrant glassy nature.

## 2. Experimental details:

Polycrystalline single phase LCMO nanoparticles of size ~ 28 nm were prepared by conventional sol-gel method. A complete details on preparation and processing conditions were reported in our previous study.[18] X-ray photoelectron spectroscopy spectra were recorded on a PHI 5000 VersaProbeII system. DC and AC magnetic susceptibility measurements were done with Quantum design SQUID-VSM magnetometer.

## 3. Results and discussion:

### 3.1 Structural study

The high resolution X-ray diffraction (HRXRD) data of LCMO nanoparticles in the Fig. 1 shows single phase pure compound. The Rietveld refinement of the experimental data using the FullProf Suite is also given in the Fig. 1; crystal structure can be assigned to pseudo tetragonal system with lattice parameters: a = b = 5.51 Å, c = 7.78 Å, α = β = γ = 90° and cell volume (V) = 236.72 (Å$^3$) with refining parameters are: $\chi^2$ =1.64, $R_p$= 10.5, $R_{wp}$ = 18.9. This pseudo tetragonal crystal structure is consistent with the low temperature synthesized LCMO powders by Dass et al.[19]

## 3.2 XPS study

The core-level X-ray photoelectron spectroscopy (XPS) of Co and Mn transition metals in LCMO nanoparticles measured at room temperature are shown in the Fig. 1(b) & (c) respectively. The 2p core-level spectrum shows Co $2p_{3/2}$ peak consisting of two peaks at the binding energies (BE) of 780.1 eV and 779.24 eV respectively. In pure cobalt oxide (CoO), Co $2p_{3/2}$ appears at 780.5 eV for Co$^{2+}$ state and in the case of Co$_3$O$_4$ presence of both 2+ and 3+ states appear at 780.5 eV and 779.6 eV respectively.[20,21] Fig.1 (c) shows the Mn 2p core-level spectrum for $2p_{3/2}$ peak in the spectrum consists of double peaks at the BE's of 642.62 eV and 641.12 respectively. The Mn-$2p_{3/2}$ peak at the BE's 642.2 eV and 641.3 eV matches with the BE's of Mn$^{4+}$ in MnO$_2$ and Mn$^{3+}$ in Mn$_2$O$_3$ compositions respectively.[21,22] This illustrates the mixed valance state of Mn and Co in present LCMO nanoparticles.

## 3.3 DC magnetization with temperature and magnetic field

Our previous magnetization study on LCMO nnaoparticles has revealed three magnetic transitions.[18] The first transition at 218 K has been attributed to the FM superexchange interactions of the ordered Co$^{2+}$ ($t_{2g}^3 e_g^2$)- O$^{2-}$- Mn$^{4+}$ ($t_{2g}^3 e_g^0$) pair and the second transition at 135

K can be assigned to FM vibronic superexchange interaction of intermediate spin $Co^{3+}$ ($t_{2g}^3 e_g^1$) – high spin $Mn^{3+}$ ($t_{2g}^3 e_g^1$) pair. Existence of these valance states are clarified from the above XPS analysis. Temperature dependent dc magnetization in zero-field-cooled ($M_{ZFC}$) and field-cooled ($M_{FC}$) process for 30 Oe dc field is shown in Fig. 2(a), where a broad transition can be observed at low temperature ~ 38 K (Fig. 2 (b)). As shown in the Fig. 2(a), a large difference between $M_{FC}$ and $M_{ZFC}$ has been observed and such a behavior is common for anisotropic FM systems and metastable magnetic glassy materials. But the bifurcation (the separation in between FC and ZFC curves) gets disappeared for applied fields above the anisotropy field in FM systems with large anisotropy. In LCMO nanoparticles, observation of bifurcation below 34 K even at 2 T (inset to Fig. 2(a)) suggests the glassy nature at low temperatures.[23] Further, $M_{FC}$ increases continuously at low temperatures below the 38 K transition. In glassy systems, a continuously increasing of $M_{FC}$ below the glassy transition in typical CG and canonical SG systems,[5,10,24] whilst in RSG systems $M_{FC}$ decreases with decreasing temperature below the glassy region.[11] We have measured thermoremanent magnetization (TRM) with temperature for 5 T field and is shown in the Fig. 2(c). This measurement was done by cooling the system from 300 K to 5 K in the presence of 5 T field and then the field was turned off and magnetization was recorded with increasing temperature. Here the TRM decreases with the increasing of temperature and shows a sudden change at ~ 218 K due to FM transition and slope variation at 38 K due to freezing of magnetic spins. Temperature variation of inverse magnetic susceptibility ($\chi^{-1}$) with a dc field of 30 Oe is shown in the Fig. 2(d). Here LCMO nanoparticles follow the Curie-Weiss (CW) law behavior in the paramagnetic (PM) region and obtained Curie - Weiss temperature ($\theta$) ~ 214 K and effective PM moment is $\mu_{eff}$ ~ 5.95 $\mu_B$. Theoretically the effective PM moment with spin only contribution can be calculated using $\mu_{eff} = \sqrt{\mu_{Co}^2 + \mu_{Mn}^2}$ formula; where $\mu_{Co/Mn} = g\sqrt{s(s+1)}$.[25] Accordingly, the $\mu_{eff}$ value is 5.44 $\mu_B$ for high spin state of $Co^{2+}$ and $Mn^{4+}$ and 6.92 $\mu_B$ for high spin state of $Co^{3+}$ and $Mn^{3+}$ states respectively. The presence of short range correlations sometimes also affects the paramagnetic moment. The variation of $\chi^{-1}$ with temperature satisfies the CW law behavior in the PM regime (Fig. 2(d)) suggests that the absence of such short range correlations above the Curie temperature. The obtained intermediate value of $\mu_{eff} \sim 5.95$ $\mu_B$ in the LCMO nanoparticles suggests the role of multiple valance states of Co and Mn as it is evident from XPS data.

The magnetic field dependence of magnetization at 5 K is shown in the Fig. 3 and obtained saturation magnetization ($M_S$) is $\sim 3.68$ $\mu_B$/f.u.; this is much smaller than the theoretically calculated spin only value of 6.0 $\mu_B$/f.u. Further, an incomplete saturation of magnetization at 5 K up to 6 T and low remnant magnetization ($M_r$) $\sim 2.2$ $\mu_B$/f.u. can be understood due to the existence of antisite disorder.[19] Inset (a) to Fig. 3 shows the variation of coercive field ($H_c$) with respect to temperature. The obtained $H_c$ values are small in the FM region and increases for temperatures below 150 K and rises sharply below 38 K. This sharp rise below 38 K suggests the random freezing of magnetic clusters.[4,24] Due to the large amount of antisite disorder one can expect that exchange bias or training effect below $T_g$ in LCMO.[9,26] We have performed field dependent isothermal (at 5 K) magnetization under 3 T FC protocol from 300 K to the measured temperature repeatedly. As shown in the inset (b) to Fig. (3), the magnetic hysteresis loops are symmetric and identical for all cycles (n= 1 to 8). Arrangement of various magnetic interactions and their strength or in other words magnetic microstructure is the key decisive factor in EB effect and the absence of training effect rules out such a scenario in LCMO.

### 3.4 DC magnetization with time

To elucidate the glassy behavior of the 38 K transition, time dependent magnetic relaxation and dynamic magnetization with temperature and dc magnetic fields are necessary. The metastable property of low-temperature glassy phase has been investigated by magnetic relaxation measurement as shown in the Fig. 4. Time dependent relaxation measurements were performed by cooling the sample from room temperature to a temperature below the glassy transition (20 K) and soaked there for different (waiting) times ($10^2$, $10^3$ and $5 \times 10^3$ sec) and then the time variation of growth in magnetization was recorded with 100 Oe dc field. Though the magnetization versus time measurement shows no complete saturation behavior but the relaxation of M (t)/M (t = 0 sec) becomes smaller with the increase of $t_w$ signify the aging phenomena. Further, it is convenient to use the relaxation rate called magnetic viscosity, defined by $S(t) = \frac{1}{H}\left(\frac{M}{\log t}\right)$ and the dependence of S (t) on $\log_{10} t$ is shown in the inset to Fig.4. The inflection point of magnetization in S (t) curve occurs close to their $t_w$ values confirm the glassy nature of the LCMO nanoparticles.

### 3.5 AC susceptibility as a function of temperature and frequency

Magnetic glassy state shows characteristic frequency dependence in the ac susceptibility. Fig. 5 shows the temperature dependent out-of-phase ($\chi''$) component of ac susceptibility for various frequencies. Two magnetic transitions were observed at 218 and 135 K are in agreement with the previous observation.[18] No frequency dependence was observed for these transitions. On the other hand the transition below 38 K exhibits strong frequency dependence as shown in the inset to Fig. 5. Such a frequency dependence of ac susceptibility at low temperature is a clear evidence of magnetic glassy state. The frequency sensitivity factor $(K) = \frac{\Delta T_f}{T_{f_0}\Delta(\log_{10}(f))}$ from the out-of phase component (($\chi''$) has been calculated and obtained a value of ~ 0.085. Mydosh [27] has categorized the **K** values by comparing various magnetic glassy systems and given as K ~ 0.005-0.01 for SGs, ~ 0.03-0.06 for CG compounds like $La_{0.5}Sr_{0.5}CoO_3$ and K>0.1 for superparamagnetic (SPM). **K** is in the range of ~ 0.06 - 0.09 has been found when magnetic nanoparticles were dispersed in a non magnetic medium.[28] The obtained **K** value (~ 0.085) in LCMO nanoparticles is higher than that of CGs and it means that the interaction among clusters is either very weak or may not be there at all.

The frequency dependent freezing temperature from the imaginary component of ac susceptibility could not be fitted to Arrhenius thermal activation and this rules out the SPM nature of the nanoparticles (figure not shown). We have also analyzed the data by using the empirical Vogel–Fulcher (VF) law and the fitting parameters are not satisfactory (figure not shown). The variation of freezing temperature ($T_f$) with relaxation time ($\tau$) can be analyzed with critical slowing down model given by power law. The obtained ac susceptibility data fits well with the power law $\tau = \tau_0(\frac{T_f - T_g}{T_g})^{-ZV}$ as shown in the inset to Fig. 5, where $\tau_0$ is microscopic spin flipping time, $T_f$ (K) is frequency dependent freezing temperature at which the maximum relaxation time ($\tau = 1/2\pi f$) of the system correspond to the measured frequency, $T_g$ is glassy freezing temperature (as f → 0 Hz and $H_{dc}$→ 0 Oe) and zv denote the critical exponent. The best fitting showed a flipping time $\tau_0$ ~ 2.4 x$10^{-5}$ sec, critical exponent (zv) ~ 3.7 $\pm$ 0.14 and glassy freezing temperature $T_g$ ~ 27 K. Though $\tau_0$ value for conventional spin glass systems is of the order of $10^{-13}$ sec,[4] the large value of $\tau_0$ (~$10^{-5}$ sec from power law)[6,8] in the present LCMO nanoparticles indicates the freezing of magnetic clusters rather than the individual atomic spins. Such a large spin flip time suggests the reentry to CG nature.

**3.6 DC magnetic field imposed AC susceptibility**

We have also performed a dc field superimposed on dynamic susceptibility. As shown in the Fig. 6, the ac susceptibility component of $\chi''(\omega, T, H)$ was plotted with temperature for different external dc fields at $\omega/2\pi$ = 523 Hz. With increasing dc magnetic field, the FM phase transition (~ 218 K) shows drastic suppression of ac susceptibility, broadening and shifting of the peak to lower temperature side. A small peak corresponding to the second FM transition (~ 135 K) was completely suppressed under the applied dc field. Contrastingly, the change in susceptibility with applied dc fields is feeble at the CG transition (~ 38 K). This is in clear contrast to typical CG behavior like in manganites,[29] a large change in ac susceptibility was found with dc bias at the glassy magnetic transition. In such systems no long range order is establish due to the competing interactions, however, with the application of dc magnetic field it favors the FM interactions and grows the cluster size; correspondingly, the CG peak temperature shifts to high temperature.[26] In the present case, the interaction among the assembly of clusters can be ruled out, as there is a small shift (~ 2 K) of the CG peak and peak shift towards low temperature side with the increase of dc bias field. Though relaxation time and the critical exponents indicate the CG nature, the dc bias on ac susceptibility suggests the non-interacting clusters and this is consistent with the higher value of frequency sensitivity factor (K). In RSG systems, the effect of dc bias on ac susceptibility near the SG phase transition is negligible.[11] RSG systems are considered to have composed of mainly two parts; one being the majority spins of the FM part and second being the SG part. Below the SG transition, the spins in SG part freezes such that their $\tau_0$ and zv satisfy spin glass characteristics and subsequently long range FM order breaks up into FM clusters.[30]

In order to shed a light on the nature of glassyness in the present sample, we have performed the memory effect. In general, memory effect is verified using ZFC protocol in dc magnetization or ac magnetization with low field and frequency so that it will not affect intrinsic nature of the system [31,32]. Usually the memory effect near the waiting temperature is observed in SG (as sharp dip) and CG (as broad dip) systems. Memory effect is absent in systems with non-interacting clusters (like SPM). In the present case, the memory effect was performed by employing the ac susceptibility (in phase component, $\chi'$) at very low frequencies ( ~ 0.3 Hz) and low ac fields (~ 0.5 Oe). As the standard protocol of the memory effect,[32] the was sample cooled from 300 K to 5 K in ZFC mode with an intermediate waiting time ($t_W$) of ~3.5 hrs at 20 K below the glassy transition and record the moment in the heating mode; this is denoted as halt magnetization ($\chi'_{halt}$). In the second time the same measurement was performed without waiting time and corresponding magnetization was recorded as reference magnetization ($\chi'_{ref}$) and in both cases constant temperature ramping rate (~3 K/min) has to be maintained. Both the curves are plotted as shown in the Fig. 6 (b) and the difference between reference and halt mode, i.e., ($\Delta\chi' = \chi'_{ref} - \chi'_{halt}$) magnetization with temperature is shown in the inset to Fig.6 (b). Here, the absence of dip in the $\Delta\chi'$ vs T curve around the halt temperature (~ 20 K) suggests the absence of collective glassy behavior in the LCMO nanoparticles.

In a perfectly ordered double perovskite LCMO, the predominate FM exchange interaction is governed by the either complete $Co^{2+} – O^{2-} – Mn^{+4}$ or $Co^{3+} – O^{2-} – Mn^{+3}$ valance species. Magnetization failing to saturate with a small $M_s$ and mixed valance of Mn (in +4 and +3) and Co (in +2 and +3) supports the existence of antisite disorder. The glassy behavior is manifested by the local antisite disorder. The AFM superexchange interactions among the antisites ($Co^{3+}$- $Co^{3+}$ and $Mn^{3+}$ - $Mn^{3+}$ mediated through the oxygen ion) weakens the average FM interactions and such an inhomogeneous magnetic order leads to multiple exchange paths within a cluster. Consequently, decrease in temperature breaks the high temperature long range FM order into cluster of competing interactions that freeze below ~ 38 K with no interaction among the clusters. In spite of having competing interactions, no training effect was observed below the glassy transition. In exchange bias systems, the horizontal shift in the hysteresis loop is possible only when the magnetic anisotropy and thickness of the pinning phase (AFM) layer is higher than that of the pinned FM phase.[33] In the present case a large magnetic anisotropy (inset to (a) to Fig. 3) of dominant FM phase hinders the EB effect.

## 4 Conclusions:

Polycrystalline single phase LCMO nanoparticles with size ~ 28 nm were prepared using conventional sol-gel method. DC magnetization and dynamic scaling of the ac susceptibility measurements and absence of memory effect suggests that low temperature magnetic transition (below ~ 38 K) is a reentry from long range FM order to a frozen state of noninteracting clusters of competing interactions. The presence of randomness due to antisite disorder with competing interactions manifests the glassy behavior in LCMO nanoparticles.

**Acknowledgments**

This work was supported by DST (SR/FTP/PS-36/2007) and the authors also acknowledge IIT Kharagpur for funding VSM SQUID magnetometer. Krishnamurthy thanks CSIR-UGC, Delhi for JRF.

**Figure captions:**

FIG. 1(a). The high resolution X-ray diffraction (HRXRD) pattern with the Rietveld analysis, XPS spectra of LCMO nanoparticles: (b) Co 2p and (c) Mn 2p core-level.

FIG. 2. (a) Temperature dependent of $M_{FC}$ and $M_{ZFC}$ for 30 Oe dc field and inset shows M (T) for 2 T; (b): close view of $M_{ZFC}$ only for 30 Oe field; (c): TRM vs. Temp (K) with 5 T field cooled and (d) Curie – Weiss law fitting for inverse susceptibility ($\chi^{-1}$) vs. Temp (K).

FIG. 3. Field dependence of magnetization at 5 K; Inset (a): Variation of coercive field $H_c$ (Oe) with temperature and inset (b): M-H loop at 5 K after field cooling is done in 3 T field for number of field cycles (n).

FIG. 4. Normalized zero-field-cooled relaxation magnetization as a function of time with 100 Oe dc field for different $t_w$ values ($10^2$, $10^3$ and $5\times10^3$ sec). Inset: Variation of magnetic viscosity i.e., relaxation rate S (t) with time in 100 Oe dc field for different $t_w$ at 20 K.

FIG. 5. Temperature dependent out-of-phase $\chi''(\omega, T, H=0)$ component of ac susceptibility for different frequencies with an ac field of 1 Oe. Inset: fitting (solid line) of spin freezing temperature ($T_f$) to the power law equation $\tau/\tau_0 = [(T_f-T_g)/T_g]^{-zv}$.

FIG. 6. (a): Out of phase magnetic ac susceptibility $\chi''(\omega, T, H)$ as a function of temperature at $\omega/2\pi$ = 523 Hz and 1 Oe ac field for different applied dc fields. (b): Temperature variation of $\chi'_{ref}$ and $\chi'_{halt}$ and inset: Absence of memory effect in $\Delta\chi$ (=$\chi'_{ref}-\chi'_{halt}$) vs T around the halting temperature.

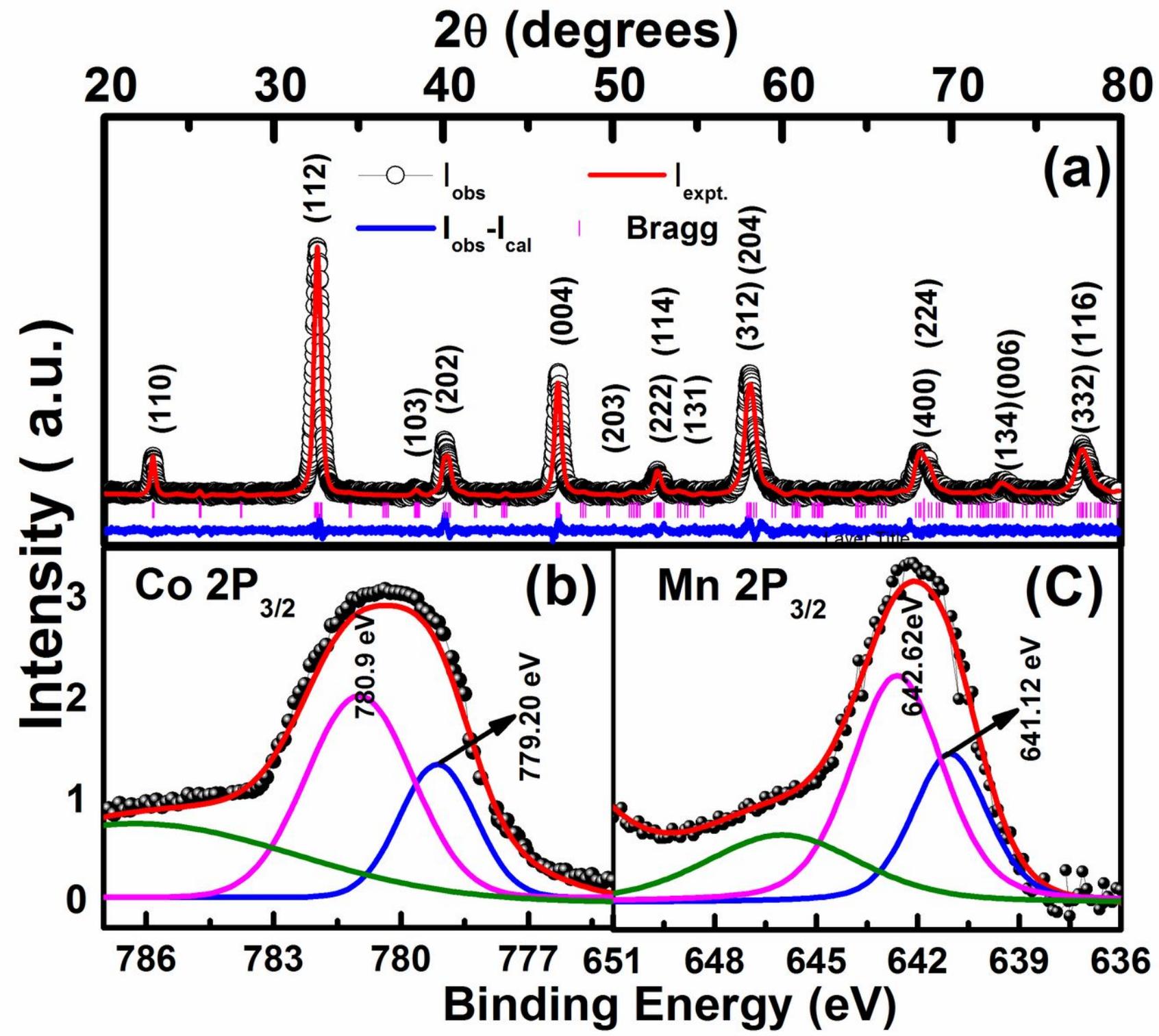

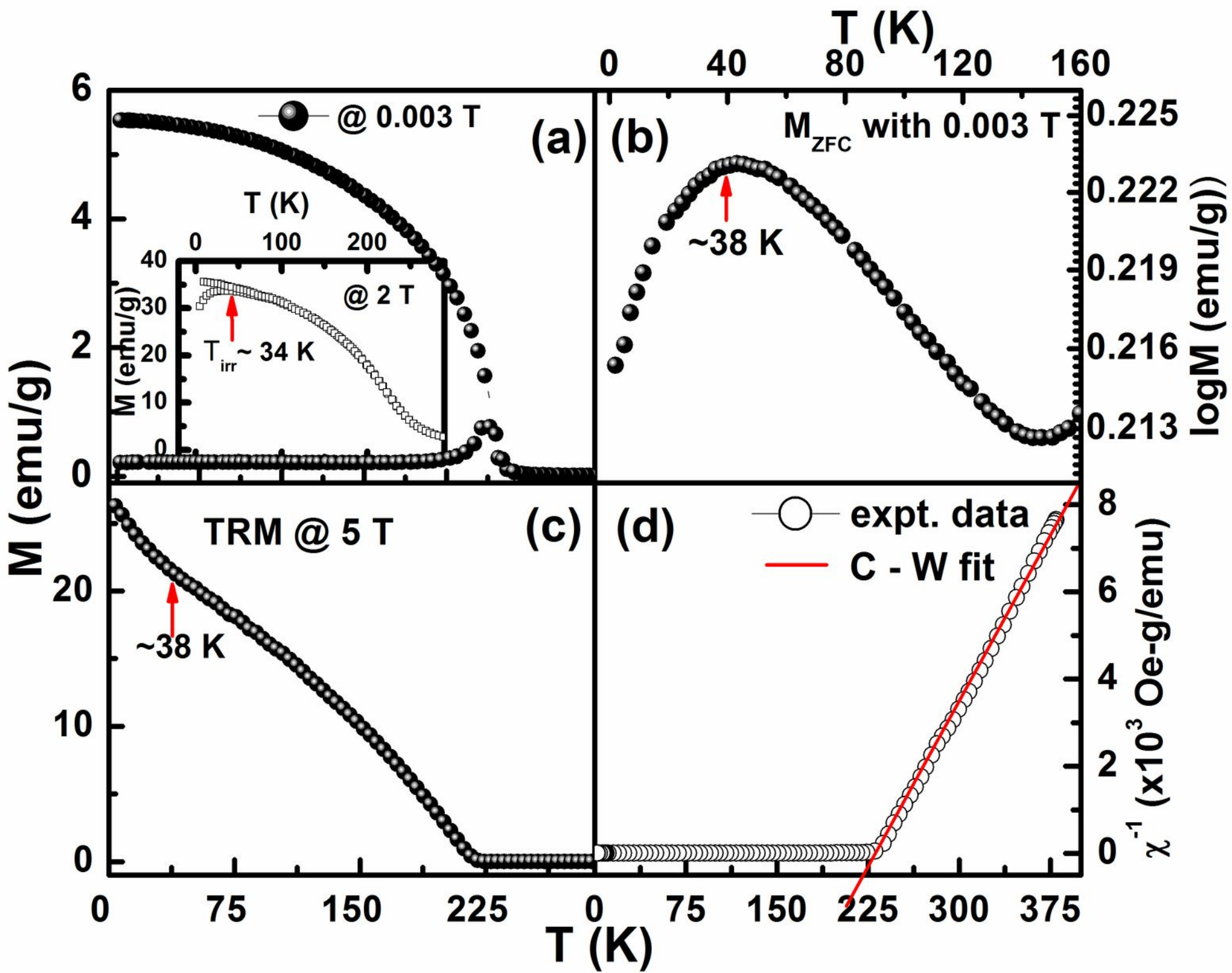

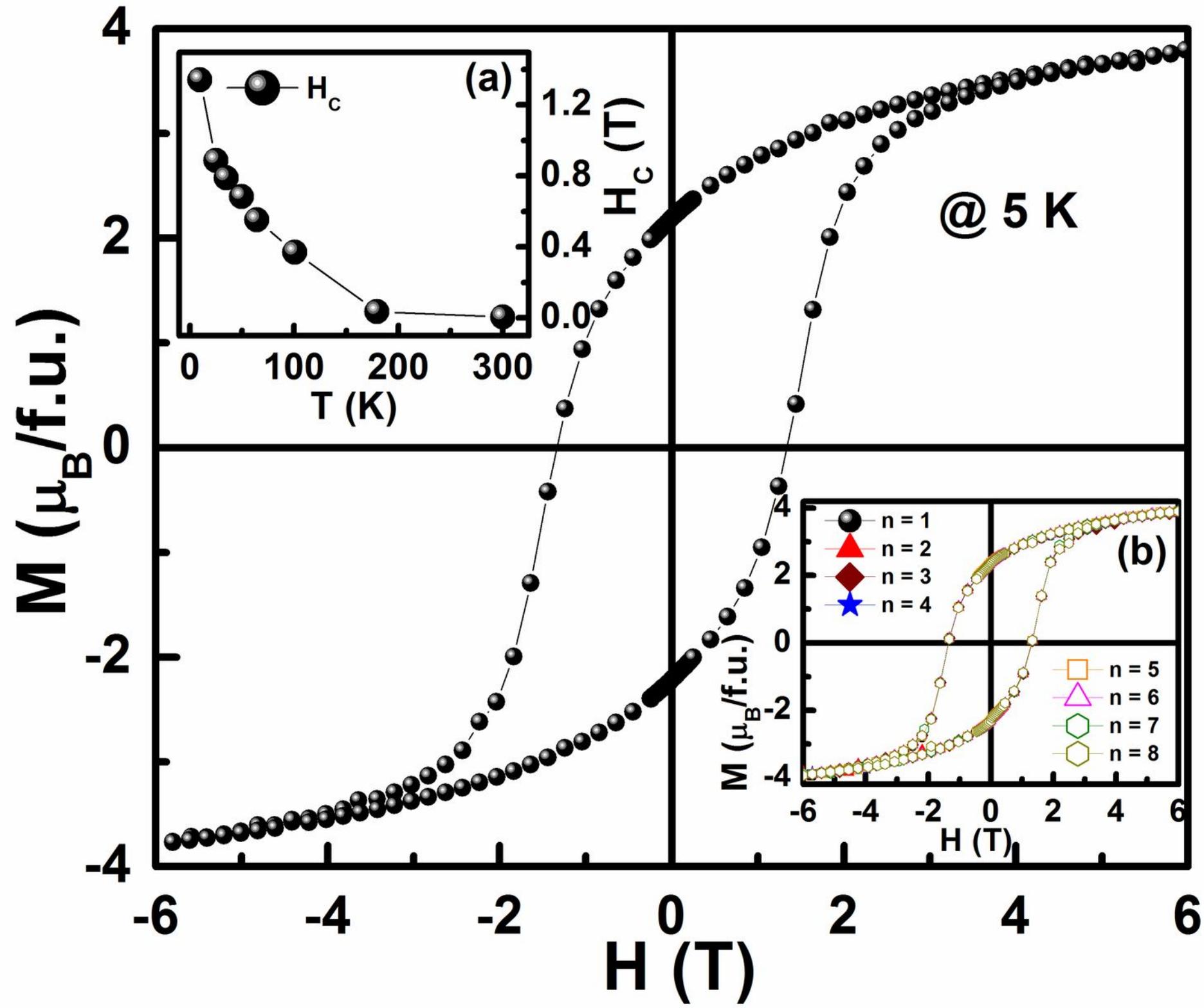

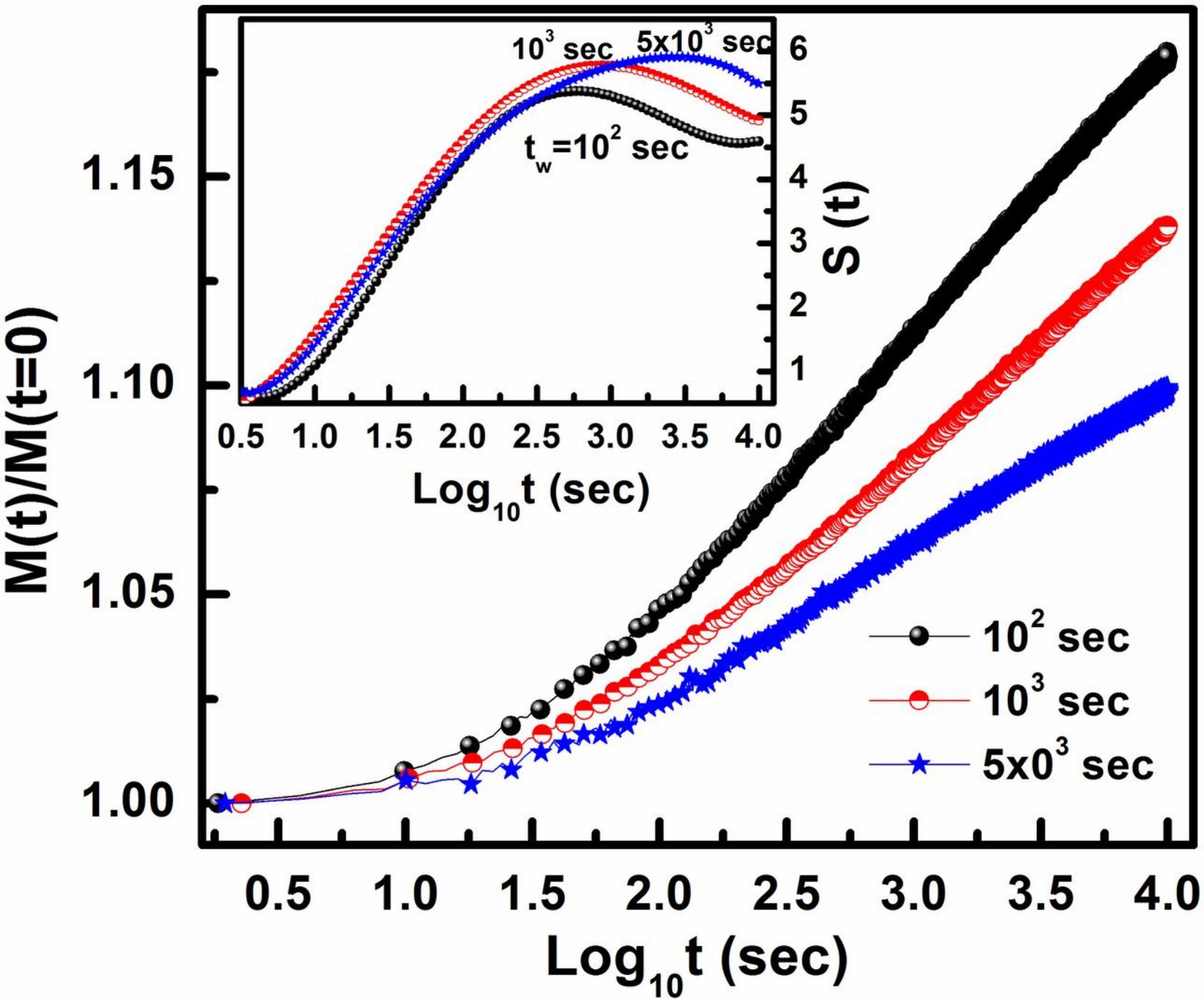

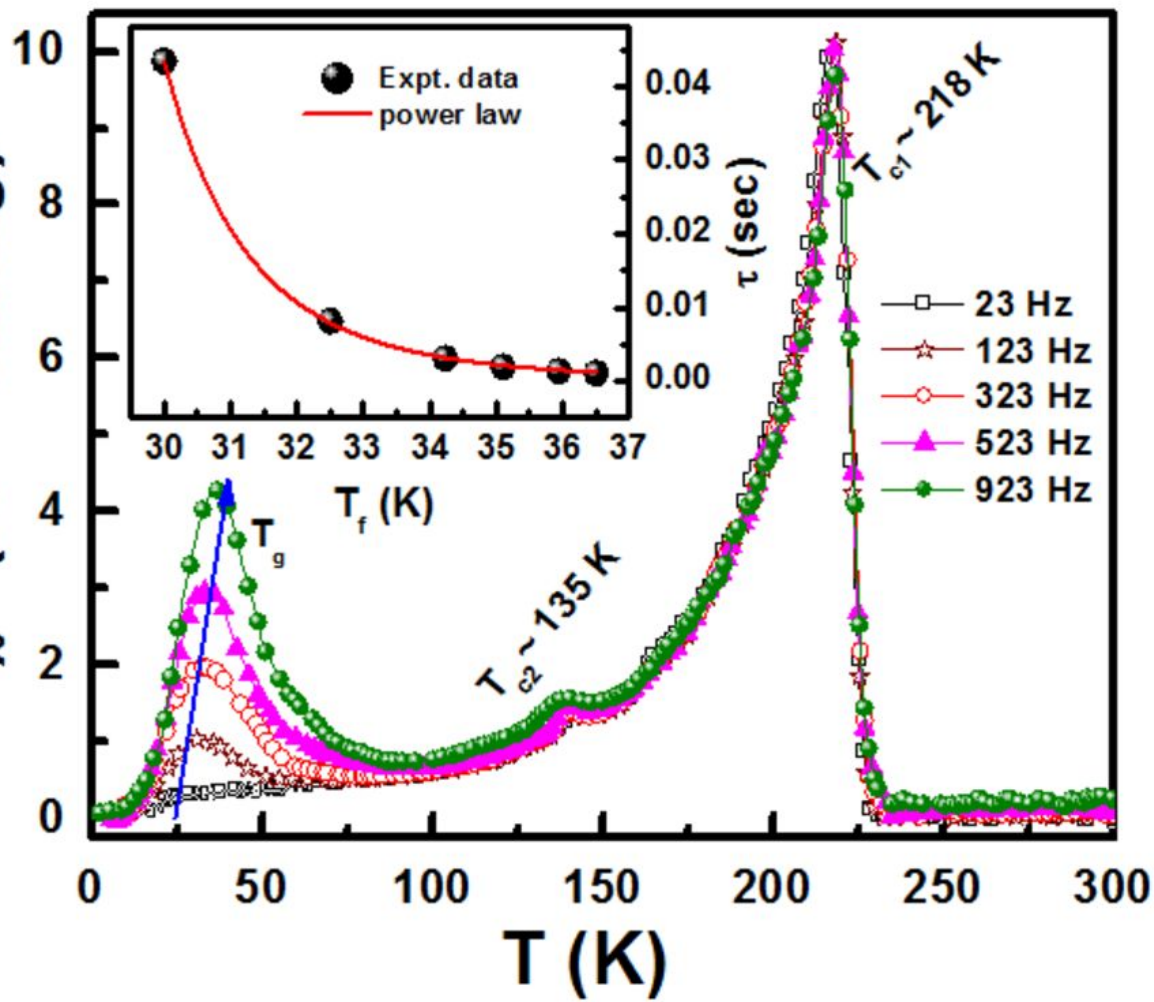

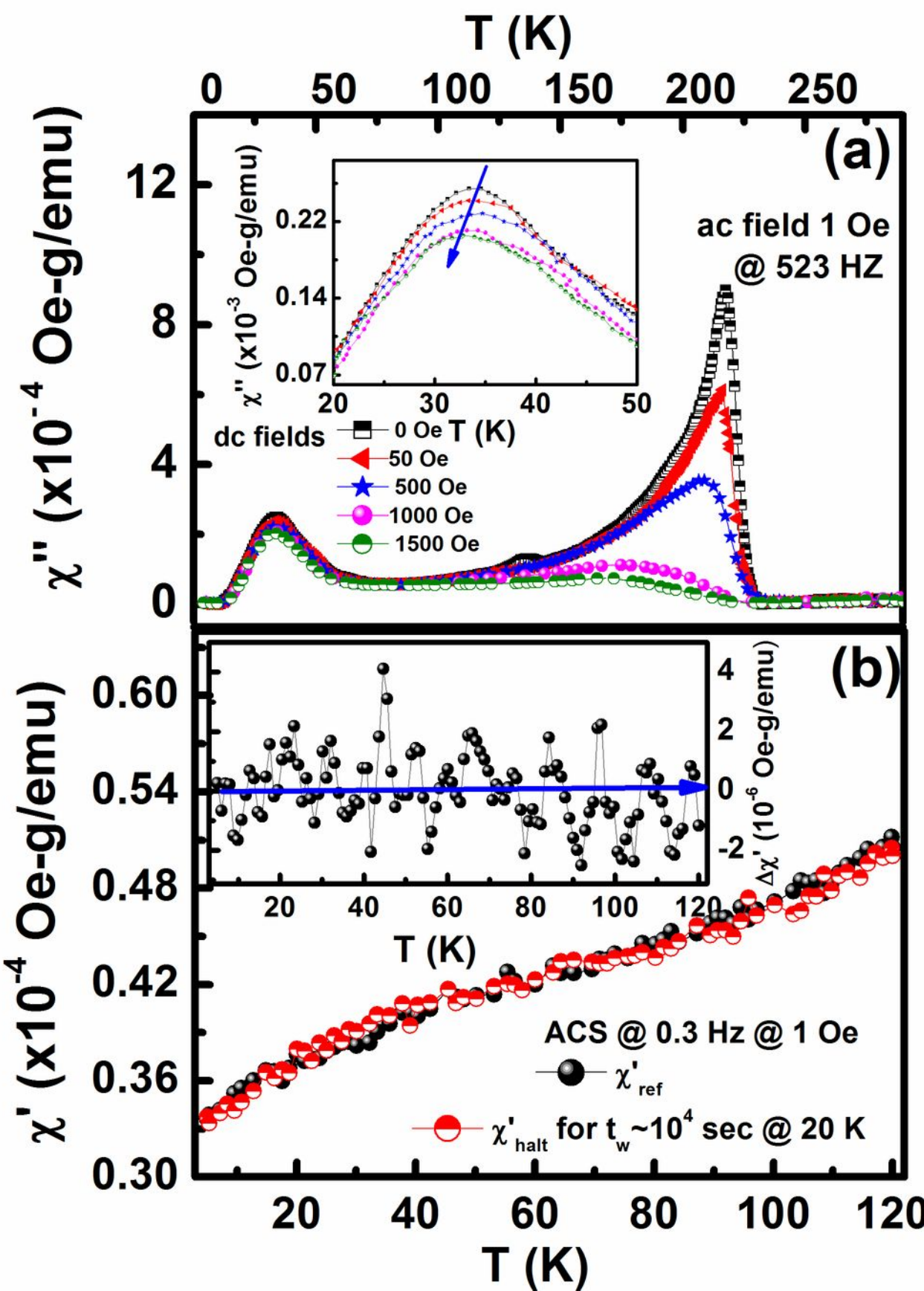